\documentclass{article}
\usepackage{spconf,amsmath,graphicx}
\usepackage{color,hyperref,enumitem,booktabs,comment,mathtools,bm}
\usepackage{multirow}
\usepackage[utf8]{inputenc}
\usepackage{balance}

\title{Joint Scattering for Automatic Chick Call Recognition}
\name{Changhong Wang\textsuperscript{1}, Emmanouil Benetos\textsuperscript{1,2}, Shuge Wang\textsuperscript{3}, Elisabetta Versace\textsuperscript{2,3}}
\address{\textsuperscript{1}Centre for Digital Music, Queen Mary University of London, UK\\
\textsuperscript{2}The Alan Turing Institute, UK\\
\textsuperscript{3}School of Biological and Behavioural Sciences, Queen Mary University of London, UK}

\begin{document}
\maketitle
\small

\begin{abstract}
Animal vocalisations contain important information about health, emotional state, and behaviour, thus can be potentially used for animal welfare monitoring.
Motivated by the spectro-temporal patterns of chick calls in the time--frequency domain, in this paper we propose an automatic system for chick call recognition using the joint time--frequency scattering transform (JTFS).
Taking full-length recordings as input, the system first extracts chick call candidates by an onset detector and silence removal. 
After computing their JTFS features, a support vector machine classifier groups each candidate into different chick call types.
Evaluating on a dataset comprising 3013 chick calls collected in laboratory conditions, the proposed recognition system using the JTFS features improves the frame- and event-based macro F-measures by 9.5\% and 11.7\%, respectively, than that of a mel-frequency cepstral coefficients baseline.
\end{abstract}

\section{Introduction}
Livestock farming is central for human sustainment.
As farming technologies are booming, the large-scale and breeding-intensive poultry industries require systems to automatically monitor the welfare of animals.
Livestock vocalisations play a crucial part in such systems, for example, assessing laying hens' thermal comfort~\cite{du2020assessment}, finding avian influenza-infected chickens to prevent the spread of diseases~\cite{cuan2020detection}, and detecting abnormal sound of broilers~\cite{liu2020novel} as an early warning tool.
Yet, to the authors' knowledge, there is limited computational research on chick vocalisations and not any prior work on automatic chick call recognition.
% or on real-time interaction between chicks and robots.
%~\cite{jakovljevic2019broiler} proposed a broiler stress detection system using a list of standard features in speech recognition. 
% System outputs were labelled at frame levels without explicitly extracting chick sound events.
%~\cite{bishop2019livestock} introduced a multi-purpose livestock vocalisation classification algorithm for classifying vocalisations of sheep, cattle, and Maremma sheepdogs in farm soundscapes. 
% The audio was manually segmented prior to classification.

Several decades ago, researchers~\cite{Joos1953spectrographic} grouped chicks' vocalisations into contact (or distress) calls and pleasure calls, analysed their characteristics through displaying them on the spectrogram, and explored the common features of the sound signals that stimulate the production of each type of call.
It was reported that contact calls are composed of descending frequencies only, are much louder, reach lower frequencies, and are given at a slower rate; while pleasure calls perform the opposite, i.e., are composed of ascending frequencies, are much softer, start from higher frequencies, and are produced at a faster rate.
Variations of chick sound patterns under successive changes in social isolation were analysed in~\cite{marx2001vocalisation}. Four types of chick calls were all labelled manually by visual inspection of the spectrogram: pleasure calls, contact calls, short peeps, and warbles.

As a pilot study towards fully automatic chick call recognition, we focus on two most frequent call types: pleasure call and contact call.
Inspecting both calls in the time--frequency domain as shown in Fig.~\ref{fig:onset_segment_detect} top, we notice that pleasure calls (0-3 s) are characterised by upward frequency changes, low energy, and short duration while contact calls (3-6 s) exhibit the opposite, i.e., downward frequency changes, high energy, and long duration, which matches the findings in~\cite{Joos1953spectrographic}.
These spectro-temporal patterns are similar to the portamento musical playing technique explored in~\cite{Wang2020jointScat}.
The latter technique also exhibits continuous frequency changes along with temporal modulations, which were shown in~\cite{Wang2020jointScat} that can be captured by the joint time--frequency scattering (JTFS).
The JTFS is an instance of the scattering transform~\cite{mallat2012group}, a signal representation that is locally invariant to translations and stable to deformations.
By adding a frequency scattering along the log-frequency axis, the JTFS exhibits extra invariance to frequency transpositions. 
These invariance properties make the JTFS a desirable representation for chick call recognition. 

% \begin{figure}[t]
%     \centering
%     \includegraphics[width=.96\linewidth]{figs/chickcall_example.png}
%     \caption{\small{Examples of pleasure (0-3 s) and contact (3-6 s) chick calls.}}
%     \label{fig:chickcallexample}
% \end{figure}

% \begin{figure}[t]
%     \centering
%     \includegraphics[width=.45\linewidth]{figs/example_pleasure.png}\hspace{.4cm}
%     \includegraphics[width=.45\linewidth]{figs/example_distress.png}
%     \caption{Examples of pleasure and distress chick calls.}
%     \label{fig:chickcallexample}
% \end{figure}

To the authors' knowledge, this is the first study to automatic recognition of chick calls using the joint time--frequency scattering transform.
We propose a recognition system comprising two stages: (1) detection stage: we extract chick call candidates using an onset detector, followed by silence removal; (2) recognition stage: with the JTFS features extracted, a support vector machine classifier groups each candidate into different chick call types. 
The results show that the JTFS features outperform a mel-frequency cepstral coefficients baseline, with the frame- and event-based macro F-measures improved by 9.5\% and 11.7\%.
We introduce each stage in Sections~\ref{sec:detection} and~\ref{sec:recognition}, respectively. Section~\ref{sec:evaluation} presents the evaluation dataset, metrics, and results, and Section~\ref{sec:conclusion} concludes the paper.

%%%%%%%%%
\section{Chick call detection}\label{sec:detection}
The number of calls emitted by the chicks is a good indicator of their state~\cite{marx2001vocalisation}.
Automatic detection of chick calls can either be a preprocessing stage for chick call recognition or as a separate system specifically optimised for chick call counting. 
We investigate the former case by firstly detecting chick call onsets and then removing silence to extract chick call segments.
We detect the onsets of chick calls using the \emph{SuperFlux}~\cite{bock2013maximum} algorithm, which
% an algorithm outperforming the benchmark spectral flux method~\cite{masri1996computer} for onset detection.
calculates the difference per frequency band in the magnitude spectrogram, applies a maximum filter along the frequency axis, sums up all positive changes over all bands, and selects the final onsets using peak-picking.
The application of the maximum filter before summing up the positive changes reduces the number of false positives originated from frequency modulations without missing onsets.
% The implementation of the onset detection algorithm is based on the Librosa Python package~\cite{mcfee2015librosa}, which takes the log-melspectrogram as input.
% We use 8 mel bands, ranging from 2048 to 6000 Hz.
% This is motivated by the observation (see Fig.~\ref{fig:onset_segment_detect}) that the energy of both pleasure and contact calls concentrates over this frequency range.
% For other parameters, we use default 
% The onset strength is calculated from the log-melspectrogram; with the size (in frequency bins) of the local max filter and the time lag for computing differences specified, i.e. specify the size (in frequency bins) of the local max filter as 18 and the time lag for computing differences as 3, respectively, based on experimental results. 
% Other parameters of the onset detection method
% The hyperparameters for the conset detection
% window = 0.15#0 # 200ms
% n_fft = 2048 * 2
% hop_length = 1024 // 2 # int(librosa.time_to_samples(1./200, sr=sr))
% lag = 3 # 2
% n_mels = 8
% fmin = 2048 # 27.5
% fmax = 6000 # 16000
% max_size = 18 # 3

\begin{figure}[ht]
    \centering
    \includegraphics[width=\linewidth]{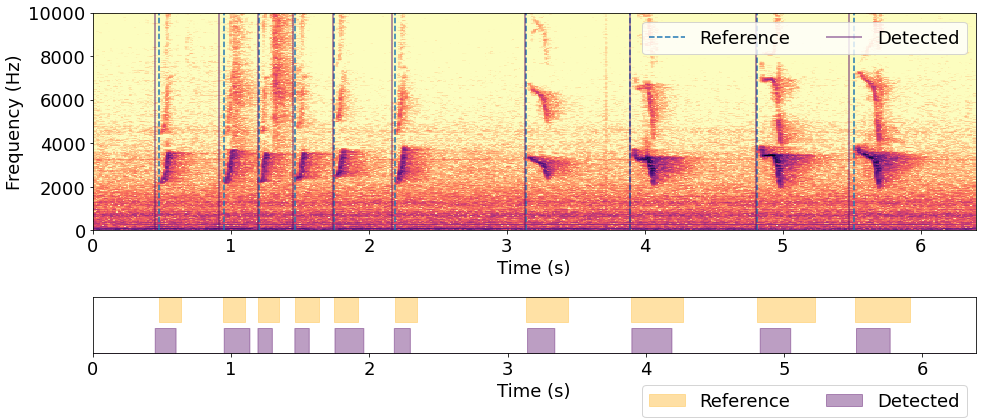}
    \caption{\small{Visualisation of onset detection and segmentation results for example chick calls. Top: spectrogram with reference onsets in dotted lines and detected onsets in solid lines; bottom: comparison of reference and detected call segments. For this example, the frame-based precision $\mathcal{P}$=96\%, recall $\mathcal{R}$=67\%, and F-measure $\mathcal{F}$=79\%.}}
    \label{fig:onset_segment_detect}
\end{figure}

Fig.~\ref{fig:onset_segment_detect} top displays the reference (dotted lines) and detected onsets (solid lines) of example chick calls.
As can be seen, the intervals between two onsets include a large proportion of silence.
We extract chick call segments from the inter-onset intervals by removing silence where the energy of the signal is below a certain threshold.
A threshold of -30 dB is used according to experimenting on a grid of threshold values.
Fig.~\ref{fig:onset_segment_detect} bottom shows the comparison of the reference call segments and the output segments of our detection system.

%%%%%%%%%
\section{Chick call Recognition}\label{sec:recognition}
% Motivated by the spectro-temporal patterns of chick calls, we calculate the JTFS on the detected segments corresponding to chick calls in Section~\ref{sec:JTFSfeature}. 
% In Section~\ref{sec:recognition}, we present the recognition system which takes the JTFS features as input and classifies each chick call candidate using a support vector machine classifier. 

% -------
\subsection{Joint Time--Frequency Scattering}\label{sec:JTFSfeature}

Similar to convolutional neural networks (CNNs), the scattering transform comprises operations of wavelet convolutions, modulus nonlinearities, and average pooling. 
Decomposing an audio waveform $\bm{x}(t)$ by a wavelet filterbank $\bm{\psi}_{\lambda}(t)$ and taking complex modulus, we obtain the first-order wavelet modulus transform, which is also known as the scalogram $\mathbf{X}(t,\lambda)$:
\begin{equation}
    \mathbf{X}(t,\lambda)  = |\bm{x} \ast \bm{\psi}_{\lambda}| (t),
\end{equation}
where $t$ and $\lambda$ are the time variable and the log-frequency variable of $\bm{\psi}_{\lambda}(t)$.
Averaging $\mathbf{X}(t,\lambda)$ by a lowpass filter $\bm{\phi}_{T}(t)$ with averaging scale $T$, the first-order scattering transform $\mathbf{S}_{1}(t,\lambda)$ is defined as 
\begin{equation}
    \mathbf{S}_{1}(t,\lambda) = \Big(|\bm{x} \ast \bm{\psi}_{\lambda}| \ast \bm{\phi}_{T} \Big)  (t).
\end{equation}

Although $\mathbf{S}_{1}(t,\lambda)$ is invariant to time shifts, the averaging comes at the detriment of fast temporal modulations with time structures smaller than $T$.
To recover these temporal modulations and to capture the variation along log-frequency axis, we decompose the scalogram with a joint time--frequency wavelet convolution, complex modulus, and averaging, according to~\cite{Anden2019joint}.
Therefore, the obtained representation has all desirable properties: stability to time warps, and invariance to time shifts and frequency transpositions.

Convolving the scalogram $\mathbf{X}({t,\lambda})$ with a spectro-temporal wavelet filterbank $\mathbf{\Psi}_{v_{\mathrm{t}},v_{\mathrm{f}},\theta}(t,\lambda)$, taking complex modulus, and averaging by a two-dimensional (2-D) lowpass filter $\mathbf{\Phi}_{T,F}(t, \lambda)$, the joint time--frequency scattering transform (JTFS)~\cite{Anden2019joint} is defined as:
\begin{equation}
    \mathbf{S}_{2}(t,\lambda, v_{\mathrm{t}}, v_{\mathrm{f}}, \theta) = \Big( \big| \mathbf{X} \overset{t,\lambda}{\ast} \mathbf{\Psi}_{v_{\mathrm{t}},v_{\mathrm{f}},\theta} \big| \overset{t,\lambda}{\ast} \mathbf{\Phi}_{T,F} \Big)(t,\lambda).
\end{equation}
$T$ and $F$ are the temporal and spectral averaging scales of $\mathbf{\Phi}_{T,F}(t,\lambda)$, respectively.
The symbol $\overset{t,\lambda}{\ast}$ denotes a 2-D convolution over the time variable $t$ and the log-frequency variable $\lambda$.
% When applied to the 2-D scalogram $\mathbf{X}(t, \lambda)$, this 1-D convolution is implicitly broadcast over the variable $\lambda$.
The 2-D joint wavelet filterbank $\mathbf{\Psi}_{v_{\mathrm{t}},v_{\mathrm{f}},\theta}(t, \lambda)$ is derived from two 1-D wavelet filterbanks, the temporal filterbank $\bm{\psi}_{v_{\mathrm{t}}}(t)$ and the spectral filterbank $\bm{\psi}_{v_{\mathrm{f}},\theta}(\lambda)$, by:
\begin{equation}\label{eq:jointWavelets}
    \mathbf{\Psi}_{v_{\mathrm{t}},v_{\mathrm{f}},\theta}(t,\lambda) =
    \bm{\psi}_{v_{\mathrm{t}}}(t)
    \bm{\psi}_{v_{\mathrm{f}},\theta}(\lambda).
\end{equation}
$v_{\mathrm{t}}$ and $v_{\mathrm{f}}$ are the log-frequency variables of $\bm{\psi}_{v_{\mathrm{t}}}(t)$ and $\bm{\psi}_{v_{\mathrm{f}},\theta}(\lambda)$, and measures the temporal and spectral variabilities, respectively.
An orientation variable $\theta=\pm 1$ is introduced to reflect the oscillation direction (up or down) of the spectro-temporal pattern. $\theta=-1$ flips the centre frequency of ${\bm{\psi}_{v_{\mathrm{f}},\theta}(\lambda)|}_{\theta=1}$ from $2^{-v_{\mathrm{f}}}$ to $-2^{-v_{\mathrm{f}}}$.

\begin{figure}[ht]
    \centering
    \includegraphics[width=.49\linewidth]{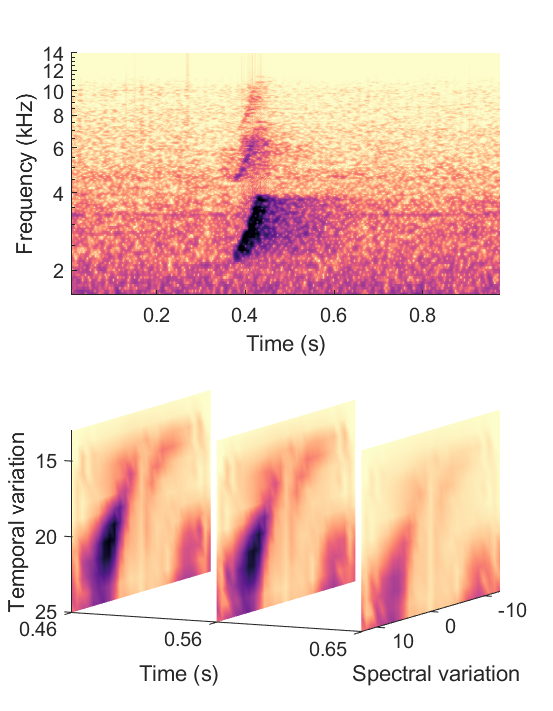}
    \includegraphics[width=.49\linewidth]{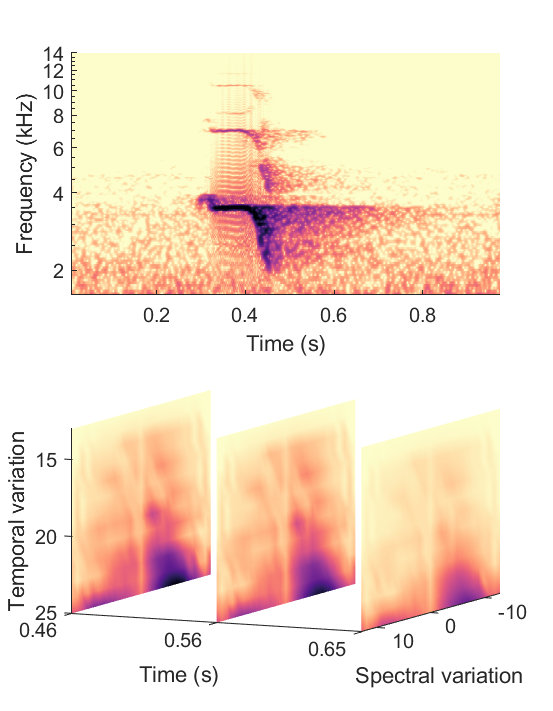}
    \caption{\small{The joint time--frequency scattering transform (JTFS) representation of example pleasure (left) and contact (right) calls. Top: log-frequency spectrogram; bottom: frame-wise JTFS features.}}
    \label{fig:JTFSofChickCalls}
\end{figure}

Fig.~\ref{fig:JTFSofChickCalls} shows the log-frequency spectrograms and the frame-wise JTFS features of a pleasure call and a contact call. 
The directions of frequency changes are captured by the clear slopes, as shown by the bottom figures.
For pleasure calls, the energy concentrates on the upward (left) side, while that of the contact call appears on the downward (right) side.
Since chick calls with upward and downward frequency changes over time belong to different classes, i.e., pleasure calls and contact calls, we use the JTFS of both directions.

\subsection{Recognition System}\label{sec:recognitionSystem}
We extract the JTFS features for chick calls by setting appropriate hyperparameters.
An averaging scale $T=2^{14}$ (in samples, corresponding to 372 ms) is used to roughly cover the average chick call duration.
Filters per octave of the temporal filters in the first-order wavelet decomposition $Q_{1}^{(\mathrm{t})}=16$ are used to offer a fine resolution scalogram.  
$Q_{2}^{(\mathrm{t})}=2$ and $Q_{1}^{(\mathrm{f})}=2$ are the filters per octave of the temporal and spectral wavelet filterbank used in the joint wavelet convolution due to the less oscillatory nature of the signal at these decompositions.
$M=[0,50]$ Hz is the extracted range of temporal modulation that contains the core information of chick calls.
$\alpha=2$ is a parameter designed to compensate for the large averaging scale $T$; it is inversely log-proportional to $T$.
To account for temporal context, we calculate the mean and standard deviation of the JTFS features of 5 frames centred at the current frame.
The frame size and dimension of the resulting JTFS features are 92 ms and 850, respectively.
With the JTFS features calculated, we propose a chick call recognition system with two classification schemes:
\begin{itemize}
    \item Scat-Only: in this scheme, a machine learning classifier takes the JTFS features of the whole recording as input and outputs frame-wise labels; neighbouring frames with the same label are then fused into chick call events. We postprocess the obtained chick call events by gap filling and minimum duration pruning. We fill the gaps between neighbouring events when the gaps are shorter than the shortest event in the training set; and prune the events that have smaller duration than the minimum duration event in the training set. The minimum duration is automatically calculated subject to the call type and the train-test split during recognition.
    \item Seg-Scat: this scheme builds upon the detection system introduced in Section~\ref{sec:detection}. A machine learning classifier takes the JTFS features of the extracted chick call segments as input, outputs frame wise labels, and assigns one label to each segment based on the majority vote of its frame labels.
\end{itemize}

Both classification schemes use support vector machines (SVMs) \cite{trevor2009elements} with Gaussian kernels as classifiers due to their good generalisability based on a limited amount of training data~\cite{albu1999application}.
In the recognition process, we split the dataset (see Section~\ref{sec:chickcalldataset}) into training and test sets by leaving one chick subject out.
% , i.e. leaving recording of one chick in the test set and those of the remaining chicks in the training set. 
Within each split, we run a 3-fold cross-validation, sampling the training set in a way that ensures approximately the same ratio of positive and negative class instances for a given chick call type in each fold.
The SVM hyperparameters to be optimized are the error penalty parameter $C$ and the width of the Gaussian kernel $\gamma$. 
We use consistent parameter grids of $10^{\{0:1:2\}}$ and $10^{\{-4:1:-2\}}$ for $C$ and $\gamma$, respectively, during training and select the best ones for testing.
For the Scat-Only scheme, we calculate the JTFS features of full-length recordings, while for the Scat-Only method, the JTFS features are calculated for the detected chick call candidates. 
All features are z-score normalised.

%%%%%%%%
\section{Evaluation}\label{sec:evaluation}
\subsection{Dataset}\label{sec:chickcalldataset}
As a pilot study\footnote{All experiments in this study were approved by the Animal Welfare and Ethical Review Body committee at Queen Mary University of London.}, we collected data in laboratory conditions.
% with a schematic shown in ref{fig:chickrecordingset}.
% The arena is an open plastic box of size $60\rm \;cm\times92\rm\;cm\times52\rm\;cm$.
% The walls of the arena are lined with white plastic, and the floor is lined with paper towels.
% We divide the arena into 5 regions---start, far, centre, close, and touch---and place the chick in the middle of the start region.
% A test stimulus (an imprinting object or a robot), is placed in the region furthest from the chick, as shown by the blue circle in the figure.
The experiments used chicks from the Ross 308 strain of the species \emph{Gallus gallus}. 
% Chicks were hatched in darkness and in individual boxes so that they had no visual or tactile experience prior to the experiment. 
We placed one chick at a time in an arena within 12 hours after their hatching.
% The chick was transferred from the hatchery to the arena using a box of size $15\rm\;cm \times15\rm \;cm\times15\rm \;cm$.
For each chick, we recorded their movements and sounds for 10 minutes by a Microsoft LifeCam Studio Webcam and an AKG P170 microphone with a Behringer U-Phoria UMC204HD audio interface.
The camera was placed approximately 1~m above the centre of the arena and the microphone 1~m above the outer wall of the arena.
All data was recorded at a sampling rate of 44.1kHz/16bits. We collected one recording for one chick at a time.

As a proof-of-concept study in this paper, we evaluate the proposed system on a dataset comprising recordings of 4 chicks. 
This is due to the available annotations we have currently: start time, end time, and call type annotations for full-length recordings of 4 chicks. 
All the annotations were created by an expert in chick behaviour from the Prepared Minds Lab at Queen Mary University of London using Sonic Visualiser~\cite{cannam2010sonic}. 
The expert was previously trained to a 91.1\% agreement level with another expert in the same task on an example chick. 
For ambiguous labelling cases, decisions were made following discussions between three experts.
Three types of chick calls were annotated: pleasure, contact, and uncertain calls. 
Uncertain calls are those calls that all the three experts were not certain about.
The total duration of the dataset is 44 minutes.
Table~\ref{tab:chickcalldataset} lists the number of each type of calls in the dataset, where the number of classes is highly imbalanced. As described in Section~\ref{sec:recognitionSystem}, we conduct a subject-independent evaluation of our recognition system, with no overlap of chick subject in the training and test sets.

\begin{table}[h]\small{
    \centering
    \begin{tabular}{p{1.2cm}p{1.2cm}p{1.2cm}p{1.4cm}p{1.2cm}}
    \toprule
   Chick & Pleasure & Contact & Uncertain &  Total\\\midrule
%  \multirow{4}{*}{\shortstack[l]{Full\\recording}} 
  1 & 119  &  315 & 9   & 443\\
  2 & 617  &  492 & 146 & 1255 \\
  3 &  47  &  682 & 60  & 789 \\
  4 &  36  &  455 & 35  & 526\\\midrule
%   prototypical & 6-12   & 40 & 40 & 0 & 80\\\midrule
  Total & 819 & 1944 & 250  & 3013 \\\bottomrule
    \end{tabular}
    \caption{\small{Number of pleasure, contact, and uncertain calls produced by each chick in the chick call dataset.}}
    \label{tab:chickcalldataset}}
\end{table}
% The calls from Chick 6-12 are the prototypical calls annotated, 5 pleasure and 5 contact for each.

% \begin{table}[h]
%     \centering\small{
%     \begin{tabular}{p{1.6cm}p{.7cm}p{1cm}p{.9cm}p{1.1cm}p{.5cm}}
%     \toprule
%   Data & Chick & Pleasure & Contact & Uncertain &  Total\\\midrule
%  \multirow{4}{*}{\shortstack[l]{Full\\recording}} 
%   & 1  & 119  &  315 & 9   & 443\\
%   & 2 & 617  &  492 & 146 & 1255 \\
%   & 3  &  47  &  682 & 60  & 789 \\
%   & 4  &  36  &  455 & 35  & 526\\\midrule
% %   prototypical & 6-12   & 40 & 40 & 0 & 80\\\midrule
%   Total & 4 & 859 & 1984 & 250  & 3093 \\\bottomrule
%     \end{tabular}
%     \caption{Number of pleasure, contact, and uncertain calls produced by each chick in the dataset. The calls from Chick 6-12 are the prototypical calls annotated, 5 pleasure and 5 contact for each.}
%     \label{tab:chickcalldataset}}
% \end{table}
% % the four chicks with call type annotations
% % chicks with stereo annotations: 21, 32, 34, 39, 41, 45, 70, 72 

% ====== Recognition results
\begin{table*}[h!]
    \centering\small{
    \begin{tabular}{p{3cm}p{2.1cm}p{1cm}p{.9cm}p{1.2cm}p{.8cm}p{.3cm}p{1cm}p{.9cm}p{1.2cm}p{.8cm}}
    \toprule
\multirow{2}{*}{\shortstack[l]{Recognition\\system Input}} & \multirow{2}{*}{\shortstack[l]{Classification\\scheme}} & \multicolumn{4}{c}{Frame-based} & &
\multicolumn{4}{c}{Event-based}\\\cmidrule{3-6}\cmidrule{8-11}
& & Pleasure & Contact & Uncertain & Macro && Pleasure & Contact & Uncertain & Macro\\\midrule
\multirow{2}{*}{Full-length recordings}
& Scat-Only & 10.5 & 79.5 & 6.3 & 32.1 && 7.0 & 29.8 & 1.4 & 12.7\\
& MFCC-Only & 4.8 & 83.8 & 7.8 & 32.1 && 2.5 & 60.9 & 8.0 & 23.8\\\midrule
\multirow{2}{*}{\shortstack[l]{Annotated segments}} & Seg-Scat & 60.3 & 95.3 & 22.0 & \textbf{59.2} & & 63.0 & 91.5 & 21.3 & \textbf{58.6}\\
& Seg-MFCC & 16.8 & 91.3 & 7.8 & 38.6 && 19.3 & 87.3 & 5.8 & 37.5\\\midrule
\multirow{2}{*}{\shortstack[l]{Detected segments}} & Seg-Scat & 17.8 & 82.0 & 15.0 & \textbf{38.3} && 19.5 & 83.1 & 14.0 & \textbf{38.9}\\
& Seg-MFCC & 4.8 & 77.0 & 4.5 & 28.8 && 3.6 & 75.0 & 2.9 & 27.2\\\bottomrule 
    \end{tabular} 
    \caption{\small{Frame-based and event-based recognition results in terms of F-measure (\%) using full-length recordings, annotated chick call segments, and detected chick call segments as inputs, respectively, for the proposed and the baseline methods: (1) Scat-Only: recognise chick calls directly from full-length recordings using the JTFS only; (2) Seg-Sat: segment audio into chick call candidates and classify each segment using the JTFS; (3) MFCC-Only: recognise chick calls directly from full-length recordings using the MFCCs only; (4) Seg-MFCCs: segment audio into chick call candidates and classify each segment using the MFCCs. `Marco' means the macro F-measure of the three types of calls.}}
    \label{tab:chickClassfResult}}
\end{table*}

\subsection{Baseline and Metrics}\label{sec:baseline_metrics}
To the authors' knowledge, there is not yet any prior work on fully automatic recognition of chick calls. 
We compare the proposed recognition system with that taking the mel-frequency cepstral coefficients (MFCCs) as input, which were commonly used for animal vocalisation analysis~\cite{bishop2019livestock, cuan2020detection, jakovljevic2019broiler}.
Similarly to the main experiments, we conduct two baseline classification schemes, i.e., MFCC-Only and Seg-MFCC.
The former calculates the MFCCs of full recordings, outputs frame labels of chick call type, and fuses neighbouring frames with the same labels into chick call events.
The latter first extracts chick call candidates using the detection method in Section~\ref{sec:detection} and then computes the MFCCs for each segment.
The frame size and dimension of the MFCC features are 25 ms and 24, respectively.
We conduct both frame- and event-based evaluation of the detection and the recognition results.
The former compares the output chick call labels with the ground truth in a frame-wise manner, while the latter compares the predicted chick call events with the ground truth by their onset and duration.
An event is considered to be correctly detected when its onset falls in a tolerance window of $\pm100$ ms around that of the reference call and its duration is more 50\% of the reference call duration.
We use precision $\mathcal{P}$, recall $\mathcal{R}$, and F-measure $\mathcal{F}$ as the metrics for each evaluation method~\cite{muller2015fundamentals}.
\subsection{Results}
To investigate the influence of the detection stage on the final recognition stage, we first present the evaluation of both onset detection and segmentation.
We evaluate the onset detection result using the \emph{mir\_eval} library~\cite{raffel2014mir_eval}, which calculates a maximum match of the reference and the detected onset times subject to a window constraint of $\pm150$ ms.
The segmentation result is evaluated using both frame- and event-based evaluation methods introduced in Section~\ref{sec:baseline_metrics}.
Table~\ref{tab:chickOnsetDetect} shows the detection and segmentation results for each chick and those on the whole dataset.
We achieve average F-measure of 89.8\% for the onset detection stage; for the segmentation result, average F-measures of 73.3\% for frame-based evaluation, and 72.3\% for event-based evaluation are obtained.

\begin{table}[h!]
\small{
    \centering
    \begin{tabular}{p{.5cm}p{.5cm}p{.2cm}p{.2cm}p{.2cm}p{.2cm}p{.2cm}p{.2cm}p{.2cm}p{.2cm}p{.2cm}}
    \toprule
    \multirow{3}{*}{Chick} & \multirow{3}{*}{\#calls} & \multicolumn{3}{c}{Onset detection} & \multicolumn{3}{c}{Segmentation} & \multicolumn{3}{c}{Segmentation} \\
     &  &  \multicolumn{3}{c}{(\emph{mir\_eval})} & \multicolumn{3}{c}{(frame-based)} & \multicolumn{3}{c}{(event-based)}\\\cmidrule{3-11}
     & &  $\mathcal{P}$ & $\mathcal{R}$ & $\mathcal{F}$ & $\mathcal{P}$ & $\mathcal{R}$ & $\mathcal{F}$& $\mathcal{P}$ & $\mathcal{R}$ & $\mathcal{F}$ \\\midrule
% 11SM\_DC & 432 & 73.4 & 99.8 & 84.6\\
% 7SM\_DC & 443 & 90.7 & 90.5 & 90.6\\
% 30VM & 626 & 68.3 & 88.5 & 77.1\\
% 9SM & 923 & 74.7 & 97.9 & 84.8\\
1 & 443 & 74 & 93 & 83 &  56 & 80 & 66 & 62 & 77 & 69\\
2 & 1255& 95 & 98 & 97 &  75 & 70 & 72 & 68 & 67 & 67\\
3 & 789 & 88 & 99 & 93 & 87 & 82 & 84 & 78 & 86 & 81\\
4 & 526 & 77 & 97 & 86 & 61 & 86 & 71 & 65 & 81 & 72\\\midrule
% 21M & 967 & 72 & 95 & 82 & 57 & 62 & 59 & 33 & 41 & 36\\
% 70M & 647 & 79 & 98 & 87 & 58 & 62 & 60 & 50 & 61 & 55\\
% 32M & 748 & 81 & 98 & 89 & 88 & 70 & 78 & 51 & 58 & 54\\
% 39M & 987 & 98 & 99 & 99 & 97 & 79 & 87 & 94 & 95 & 94  \\
% 34F & 707 & 88 & 97 & 92 & 45 & 46 & 46 & 46 & 47 & 47\\
% 41F & 1052& 92 & 93 & 93 & 78 & 76 & 77 & 80 & 80 & 80\\
% 48F & 294 & 86 & 99 & 92 & 46 & 36 & 40 & 31 & 35 & 33\\
% 72F & 186 & 81 & 98 & 89 & 35 & 48 & 40 & 42 & 47 & 44\\\midrule
T/A & 3013 & 83.5 & 96.8 & 89.8 & 69.8 & 79.5 & 73.3 &  68.3 & 77.8 & 72.3\\
\bottomrule
    \end{tabular}
    \caption{\small{Onset detection and segmentation results for each chick in the chick call dataset ($\mathcal{P}$, $\mathcal{R}$, and $\mathcal{F}$ are in \%; T/A=total or average).}}
    \label{tab:chickOnsetDetect}}
\end{table}

Table~\ref{tab:chickClassfResult} displays the recognition results on the recordings of the four chicks.
We compare these results from four fronts: the JTFS features versus the MFCCs; the two classification schemes using the JTFS; the recognition results using detected versus annotated chick call segments; and the results of different call types.
Comparing the Scat-Only with the MFCC-Only method, we observe that they achieve comparable results in the frame-based evaluation, both with macro F-measures of 32.1\%; while in the event-based evaluation, the former underperforms the latter, with macro F-measure of 12.7\% against 23.8\%.
For the comparison between the JTFS and the MFCCs using detected segments, the Seg-Scat outperforms Seg-MFCC, with macro F-measures improved by 9.5\% in the frame-based evaluation and by 11.7\% in the event-based evaluation, respectively.
Narrowing the scope with the recognition results using the JTFS, i.e., Scat-Only and Seg-Scat, the former underperforms the latter in both frame- and event-based evaluation.
The Seg-Scat increases macro F-measure by 6.2\% as compared to that of the Scat-Only in the frame-based evaluation and by 26.2\% in the event-based evaluation.
It is expected that the Scat-Only obtains much lower event-based macro F-measures in contrast to the Seg-Scat, where the latter uses chick call segments, either detected or annotated.

As shown in Table~\ref{tab:chickClassfResult}, both `Seg-' methods, i.e., Seg-Scat and Seg-MFCC, exhibit much better performance using the annotated chick call segments as compared to using the detected chick call segments.
In the frame-based evaluation, macro F-measures increase by 20.9\% and 9.8\% for the two methods, respectively; the corresponding improvement in the event-based evaluation are 19.7\% and 10.3\%, respectively.
This verifies the potential of the JTFS for chick call recognition.
Inspecting the F-measures of each type of calls, we can see that all four methods exhibit better performance on contact call classification than that on pleasure and uncertain call classification.
This may be attributed to the small amount of samples for the latter two types of calls or to their features (e.g., lower amplitude and shorter duration).
Yet, the JTFS is more robust to data imbalance as compare to the MFCCs.
For example, the Seg-Scat achieves frame-based F-measure of 60.3\% against 16.8\% from the Seg-MFCC for recognising pleasures call using annotated segments, although both methods have comparable results on contact call classification.

\section{Conclusion}\label{sec:conclusion}
In this paper, we have proposed a fully automatic system for chick call recognition using the joint time--frequency scattering (JTFS).
The system first extracts chick call candidates from full-length recordings by onset detection and silence removal; and then classifies each candidate based on their JTFS features.
Compared with a mel-frequency cepstral coefficients baseline, the proposed system using the JTFS of detected chick call segments improves macro F-measures by 9.5\% and 11.7\% in the frame- and event-based evaluation and is more robust to data imbalance.
Comparison of the recognition performance using detected and annotated chick call segments verifies the potential of the JTFS for chick call recognition.

Three limitations exist in the current study. 
We collected data in a laboratory which is much cleaner than that recorded in real living conditions of chicks.
For the latter case, we could process the recordings with  source separation techniques to separate the sound of multiple chicks or a denoising method to remove noise, prior to the application of the JTFS features.
% We place one chick at a time in the experiment arena while in a practical cases there may be many chicks vocalising simultaneously.
% For example, a pleasure call produced by one chick may overlap with a contact call produced by another chick.
The other two limitations include the small amount of data, in terms of both duration and number of chicks, and the highly imbalanced number of examples for each class.
Lack of data is the main reason of not having applied deep learning techniques in this paper.
For future work, we may use semi-supervised or unsupervised learning for the task due to the availability of unlabelled data.
Unsupervised learning may also potentially discover new chick call patterns.
We can also compare the proposed system to vocalisation detection systems developed for other animals, for example, the broiler stress detection system in~\cite{jakovljevic2019broiler}.

% \section{Possible ways to improve the paper}
% Currently I only present the result of the four chicks with call type annotations without taking the 80 ``prototypical'' calls into account. If we want to include these calls, there may be two possible ways (the first two bullet points below).
% \begin{itemize}
%     \item Keep the whole pipeline of the current system but add these prototypical calls as extra training data. Therefore, in each split of the data, we will have 3 chicks + 80 prototypical calls in the training set, and 1 chick in the test set.
%     \item Since Eli emphasised that these prototypical calls can be regarded as template calls, I am wondering if we could use a kind of ``template-matching'' method: use the prototypical calls only as the training data to label each detected chick call candidates in all the four recordings.
%     \item Besides how to use the the prototypical calls, do you recommend to include also other baseline methods or features apart from MFCCs?
% \end{itemize}
\balance
\bibliographystyle{IEEEbib}
\bibliography{bibliography}

\end{document}